\documentclass[fleqn,usenatbib]{mnras}

\usepackage{newtxtext,newtxmath}

\usepackage[T1]{fontenc}

\usepackage{graphicx}
\usepackage{amsmath}

\newcommand*\samethanks[1][\value{footnote}]{\footnotemark[#1]}

\title[Massive quiescent galaxies at $3 < z < 5$]{A surprising abundance of massive quiescent galaxies at $\mathbf{3 < z < 5}$ in the first data from JWST CEERS}

\author[A. C. Carnall et al.]
{A. C. Carnall$^{1}$\thanks{E-mail: adamc@roe.ac.uk},
D. J. McLeod$^{1}$,
R. J. McLure$^{1}$,
J. S. Dunlop$^{1}$,
R. Begley$^{1}$,
F. Cullen$^{1}$,
C. T. Donnan$^{1}$,
\newauthor
M. L. Hamadouche$^{1}$,
S. M. Jewell$^{1}$,
E. W. Jones$^{1}$,
C. L. Pollock$^{1}$,
V. Wild$^{2}$
\medskip\\
$^{1}$ SUPA\thanks{Scottish Universities Physics Alliance}, Institute for Astronomy, University of Edinburgh, Royal Observatory, Edinburgh EH9 3HJ, UK \\
$^{2}$ SUPA\samethanks, School of Physics and Astronomy, University of St Andrews, North Haugh, St Andrews KY16 9SS, UK}
\date{Accepted XXX. Received YYY; in original form ZZZ}
\pubyear{2022}

\begin{document}
\label{firstpage}
\pagerange{\pageref{firstpage}--\pageref{lastpage}}
\maketitle

\begin{abstract}

\noindent We report a robust sample of 10 massive quiescent galaxies at redshift, $z > 3$, selected using the first data from the JWST CEERS programme. Three of these galaxies are at $4 < z < 5$, constituting the best evidence to date for quiescent galaxies significantly before $z=4$. These extreme galaxies have stellar masses in the range log$_{10}(M_*/$M$_\odot) = 10.1-11.1$, and formed the bulk of their mass around $z \simeq 10$, with two objects having star-formation histories that suggest they had already reached log$_{10}(M_*/$M$_\odot) > 10$ by $z\gtrsim8$. We report number densities for our sample, demonstrating that, based on the small area of JWST imaging so far available, previous work appears to have underestimated the number of quiescent galaxies at $3 < z < 4$ by a factor of $3-5$, due to a lack of ultra-deep imaging data at $\lambda>2\,\mu$m. This result deepens the existing tension between observations and theoretical models, which already struggle to reproduce previous estimates of $z>3$ quiescent galaxy number densities. Upcoming wider-area JWST imaging surveys will provide larger samples of such galaxies and more-robust number densities, as well as providing opportunities to search for quiescent galaxies at $z>5$. The galaxies we report are excellent potential targets for JWST NIRSpec spectroscopy, which will be required to understand in detail their physical properties, providing deeper insights into the processes responsible for forming massive galaxies and quenching star formation during the first billion years.
\end{abstract}


\section{Introduction}

Two of the most important outstanding questions in galaxy evolution are: when did the first galaxies begin to form stars, and when did the first galaxies quench their star-formation activity? During the short time since the first data from the \textit{James Webb Space Telescope} (JWST) were released, remarkable progress has been made towards addressing the first of these questions. We now have good evidence that log$_{10}(M_*/$M$_\odot) \simeq 8 - 9$ galaxies were already in place by $z\simeq17$, less than 250 Myr after the Big Bang, with preliminary evidence mounting that such objects are more numerous than expected (e.g. \citealt{Naidu2022, Castellano2022, Donnan2022, Finkelstein2022}). We have also uncovered the unexpectedly rapid growth of these early seeds into massive galaxies with log$_{10}(M_*/$M$_\odot)\simeq 10-11$ during the latter half of the first billion years, from $6 < z < 10$ \citep{Labbe2022}.

This extremely rapid assembly of the first massive galaxies is critically important for our understanding of quenching. If log$_{10}(M_*/$M$_\odot)\simeq 11$ galaxies already exist by $6 < z < 10$, these must equally rapidly quench, and remain quenched, to avoid becoming too massive to be accommodated by the lower-redshift galaxy stellar-mass function (e.g. \citealt{McLeod2021}). This suggests massive quiescent galaxies at least as early as $z\simeq6$.

Currently, the earliest spectroscopically confirmed massive quiescent galaxies are half a billion years later, when the Universe was $\simeq1.5$ Gyr old at $z\simeq4$ (e.g. \citealt{Glazebrook2017, Valentino2020, Forrest2020}), and indeed it has proven extremely challenging to identify even robust photometric candidates at $z>4$ (e.g. \citealt{Merlin2018, Merlin2019, Carnall2020, Esdaile2021, Santini2021, Stevans2021, Marsan2022}). It is currently unclear whether this is due to an almost total lack of quiescent galaxies at earlier times, or due to a lack of ultra-deep, high-resolution imaging at $\lambda > 2\,\mu$m with which to constrain the Balmer break at these redshifts.

The number density and passive fraction of high-redshift massive galaxies are, however, key constraints on galaxy formation models, with current simulations unable to reproduce the observed number density of quiescent galaxies at $3 <z<4$ (e.g. \citealt{Schreiber2018, Cecchi2019, Girelli2019}). This implies that key physics, capable of giving rise to extremely rapid quenching events, is still missing from these simulations. In this context, more-robust constraints on the number density of massive quiescent galaxies at $3 < z < 4$, and confirmation of whether any such objects exist at $z>4$, are key to our understanding of galaxy formation.

\begin{figure*}
	\includegraphics[width=0.98\textwidth]{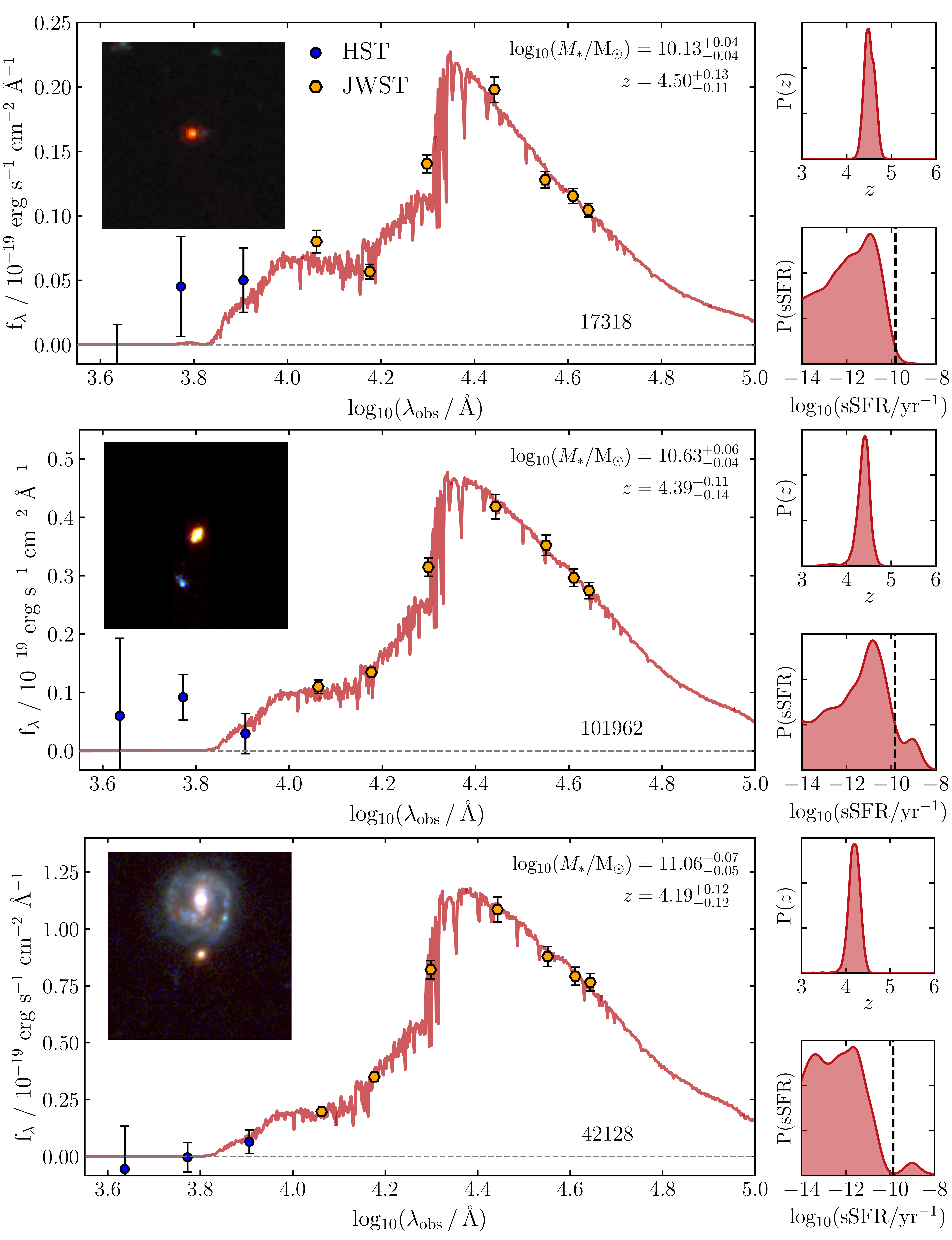}\vspace{-0.4cm}
    \caption{Spectral energy distributions and cutout images for our three robust $z > 4$ quiescent galaxies. Our 10-band photometric data from HST ACS and JWST NIRCam are shown in blue and gold respectively. The posterior median \textsc{Bagpipes} models are overlaid in red. Posterior distributions for the redshifts and sSFRs of these galaxies are shown to the right of the main panels. The dashed vertical lines in the sSFR panels show the sSFR threshold for inclusion in our quiescent sample at the redshift of each object (see Section \ref{method:sel}). The inset RGB cutouts are composed of the F444W, F200W and F150W images respectively.}
    \label{fig:spectra}
\end{figure*}

\begin{table*}
  \caption{The 9 free parameters of the \textsc{Bagpipes} model we fit to our photometric data (see Section \ref{method}), along with their associated prior distributions. The upper limit on $\tau$, $t_\mathrm{obs}$, is the age of the Universe as a function of redshift. Logarithmic priors are all applied in base ten. For the Gaussian prior on $\delta$, the mean is $\mu$ and the standard deviation is $\sigma$.}
\begingroup
\setlength{\tabcolsep}{9pt} 
\renewcommand{\arraystretch}{1.2} 
\begin{tabular}{lllllll}
\hline
Component & Parameter & Symbol / Unit & Range & Prior & \multicolumn{2}{l}{Hyper-parameters} \\
\hline
General & Redshift & $z$ & (0, 20) & Uniform & & \\
& Total stellar mass formed & $M_*\ /\ \mathrm{M_\odot}$ & (1, $10^{13}$) &Logarithmic & & \\
& Stellar and gas-phase metallicities & $Z\ /\ \mathrm{Z_\odot}$ & (0.2, 2.5) &Logarithmic & & \\
\hline
Star-formation history & Double-power-law falling slope & $\alpha$ & (0.01, 1000) & Logarithmic & & \\
& Double-power-law rising slope & $\beta$ & (0.01, 1000) & Logarithmic & & \\
& Double power law turnover time & $\tau$ / Gyr & (0.1, $t_\mathrm{obs}$) & Uniform & & \\
\hline
Dust attenuation & $V-$band attenuation & $A_V$ / mag & (0, 8) & Uniform & & \\
& Deviation from \cite{Calzetti2000} slope & $\delta$ & ($-0.3$, 0.3) & Gaussian & $\mu = 0$ & $\sigma$ = 0.1 \\
& Strength of 2175\AA\ bump & $B$ & (0, 5) & Uniform & & \\
\hline
\end{tabular}
\endgroup
\label{table:param}
\end{table*}

By providing unprecedentedly deep infrared imaging at $\lambda > 2\,\mu$m, JWST opens up a unique opportunity to address this issue, promising the ability to robustly select representative samples of massive galaxies as far back as just a few hundred Myr after the Big Bang. In addition, its extremely high angular resolution (e.g. \citealt{Suess2022}) and wide-ranging spectroscopic capabilities (e.g. \citealt{Carnall2023a}) hold much promise for extending detailed studies of quiescent galaxy physical properties, such as star-formation histories (SFHs), stellar metallicities and sizes, back to the first billion years. This endeavour has previously proven extremely challenging, even at cosmic noon (e.g. \citealt{Wu2018b, Belli2019, Carnall2019b, Carnall2022b, Beverage2021, Hamadouche2022}).

In this paper, we use extremely deep $\lambda=1-5\,\mu$m NIRCam imaging from the JWST Cosmic Evolution Early Release Science (CEERS\footnote{\url{https://ceers.github.io}}) programme (Finkelstein et al. in prep.) to search for massive galaxies at $z>3$. In particular, we focus on constraining the number density of galaxies that have already quenched their star-formation activity at this early time. We also make a first attempt at measuring the SFHs of these galaxies, despite the significant challenge of measuring these from photometric data alone, in order to link them with the extreme population of star-forming galaxies currently being uncovered during the first billion years. The spectroscopic capabilities of JWST mean that spectroscopic redshifts, SFR measurements, and even spectroscopic SFH and stellar metallicity determinations are a realistic prospect for these galaxies on a short timescale (e.g., \citealt{Carnall2023c}).

The structure of this paper is as follows. In Section \ref{data} we introduce the CEERS and ancillary datasets used in this work. In Section \ref{method} we discuss our spectral energy distribution (SED) fitting methodology and sample selection. In Section \ref{results} we present our results, including the discovery of three robustly identified massive quiescent galaxies at $4 < z < 5$. We discuss our results in Section \ref{discussion} and present our conclusions in Section \ref{conclusion}. All magnitudes are quoted in the AB system. For cosmological calculations, we adopt $\Omega_M = 0.3$, $\Omega_\Lambda = 0.7$ and $H_0$ = 70 $\mathrm{km\ s^{-1}\ Mpc^{-1}}$. We assume a \cite{Kroupa2001} initial mass function, and assume the Solar abundances of \cite{Asplund2009}, such that $\mathrm{Z_\odot} = 0.0142$.

\begin{figure*}
	\includegraphics[width=\textwidth]{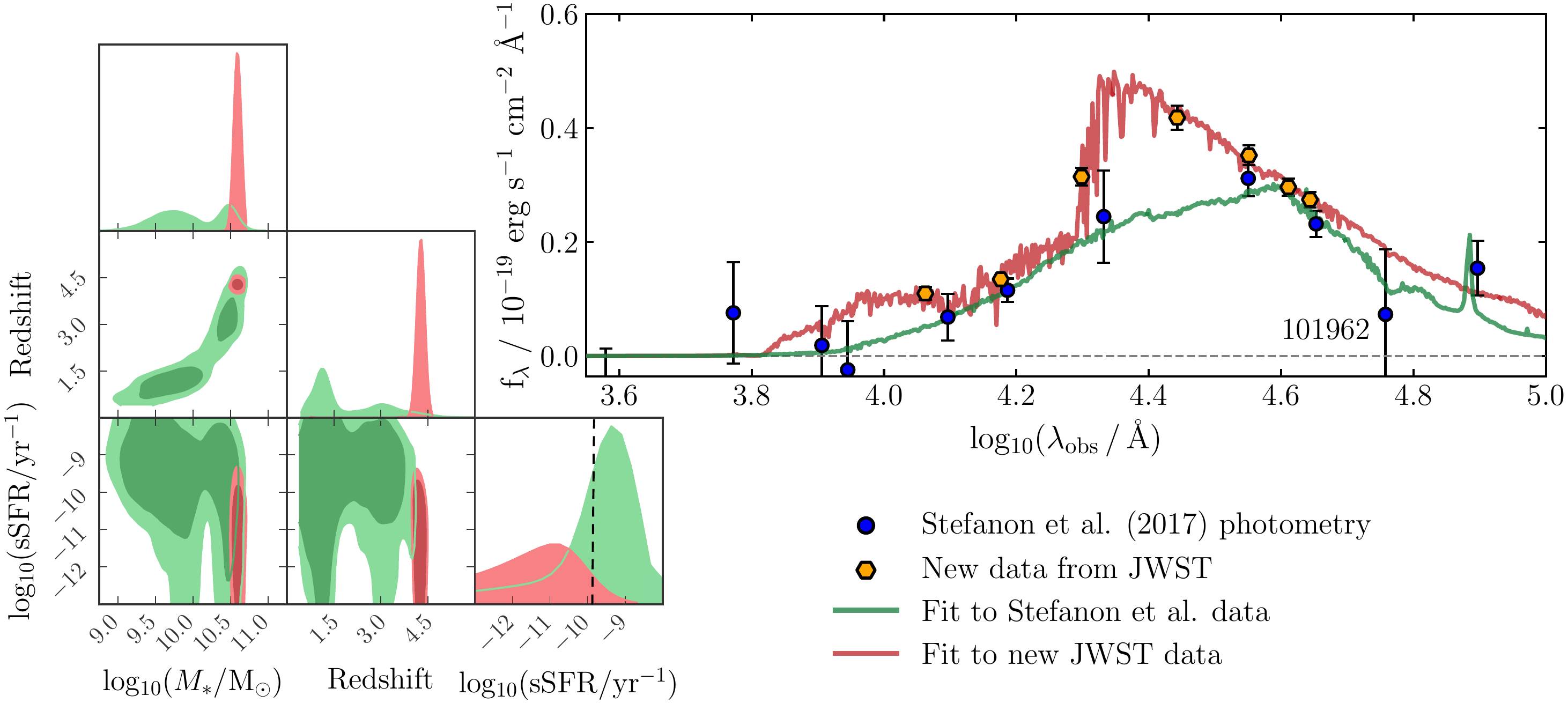}\vspace{-0.3cm}
    \caption{Comparison of pre-JWST and new JWST CEERS fit results for object 101962, the only robust $z>4$ quiescent galaxy in our sample to be included in the CANDELS photometric catalogue of \protect \cite{Stefanon2017}. The best fit to the new JWST NIRCam data (gold) is shown in red, as in the middle panel of Fig. \ref{fig:spectra}. The best fit to the CANDELS data (blue) is shown in green. The corner plot to the left shows constraints on the stellar mass, redshift and sSFR of this galaxy from both datasets. It can be seen that previous data were unable to constrain these parameters, with an extremely large redshift uncertainty, $z=1.34^{+1.94}_{-0.39}$. This is largely due to the low SNR of previous data, in particular around the Balmer break, and a lack of data at $\lambda\simeq2.3-3.0\mu$m between $K_s$ and IRAC Channel 1.}
    \label{fig:spectra2}
\end{figure*}

\section{Data}\label{data}

The primary dataset for this work is comprised of the first observations made as part of the CEERS survey (Finkelstein et al. in prep) in the CANDELS Extended Groth Strip (EGS) field. We use data from the first 4 pointings observed in late June 2022. Imaging is available in 7 NIRCam filters: F115W, F150W, F200W, F277W, F356W, F410M and F444W, with integration times of 2635 seconds per filter, except F115W where the exposure time is doubled. The currently available CEERS data amounts to a total effective area of $\simeq30$ square arcmin (e.g. \citealt{Donnan2022}). 

In addition to the NIRCam imaging, we also make use of \textit{Hubble Space Telescope} (HST) ACS data in the F435W, F606W and F814W bands. We use the v1.9 EGS mosaics produced by the CEERS team \citep{Koekemoer2011}, which have a pixel scale of $0.03^{\prime\prime}$.

We perform our own custom reduction of the NIRCam data, beginning with the Level 1 data products, using the PRIMER Enhanced NIRCam Image Processing Library (PENCIL), a custom version of the JWST pipeline (v1.6.2). We align and stack the individual reduced images using the \textsc{Scamp} and \textsc{Swarp} codes \citep{Bertin2006, Bertin2010}. Our final mosaic images have a pixel scale of $0.03^{\prime\prime}$ in all bands. We PSF-homogenise all data to the F444W band using empirical PSFs derived from stacks of bright stars in our mosaic images. We make use of the CRDS\_CTX = jwst\_0942.pmap version of the JWST calibration files, released on 28\textsuperscript{th} July 2022. We then apply the empirically derived, module-dependent calibration correction factors described in Appendix C of \cite{Donnan2022}. These calibrations are in good agreement with empirical calibrations derived by other teams\footnote{e.g. https://github.com/gbrammer/grizli/pull/107}.

We employ the \textsc{SExtractor} code \citep{Bertin1996} to measure object photometry. We run \textsc{SExtractor} in dual image mode, with the F200W mosaic used as the detection image in all cases, as this is the wavelength range in which the SEDs of $z\simeq3-5$ quiescent galaxies peak. Fluxes are measured within $0.5^{\prime\prime}$ (16 pixel) diameter circular apertures. We then aperture-correct by scaling to measurements of \textsc{FLUX\_AUTO} in the F200W band. This size of aperture is more than sufficient for our purposes, given that massive quiescent galaxies at high redshift are known to be both extremely compact and centrally concentrated (e.g. \citealt{vanderwel2014}). 

We cut our catalogue at F200W = 26.5, at which magnitude objects have a typical F200W SNR $\simeq 20$. This is necessary to ensure objects are reliably detected with high SNR in the other relevant bands, in particular to obtain strong constraints on the Balmer break strength. This results in a parent sample of 10542 objects.

We measure uncertainties for each object as the standard deviation of fluxes measured in the closest $\simeq100$ blank sky apertures \citep{McLeod2016}, whilst masking out nearby objects, using the robust median absolute deviation (MAD) estimator. We check the resulting photometry in the F606W and F814W bands by cross-matching with the CANDELS catalogue produced by \cite{Stefanon2017}, finding good agreement (e.g., median offset of 0.07 magnitudes in F814W, with a scatter of 0.14 magnitudes).

\begin{figure*}
	\includegraphics[width=\textwidth]{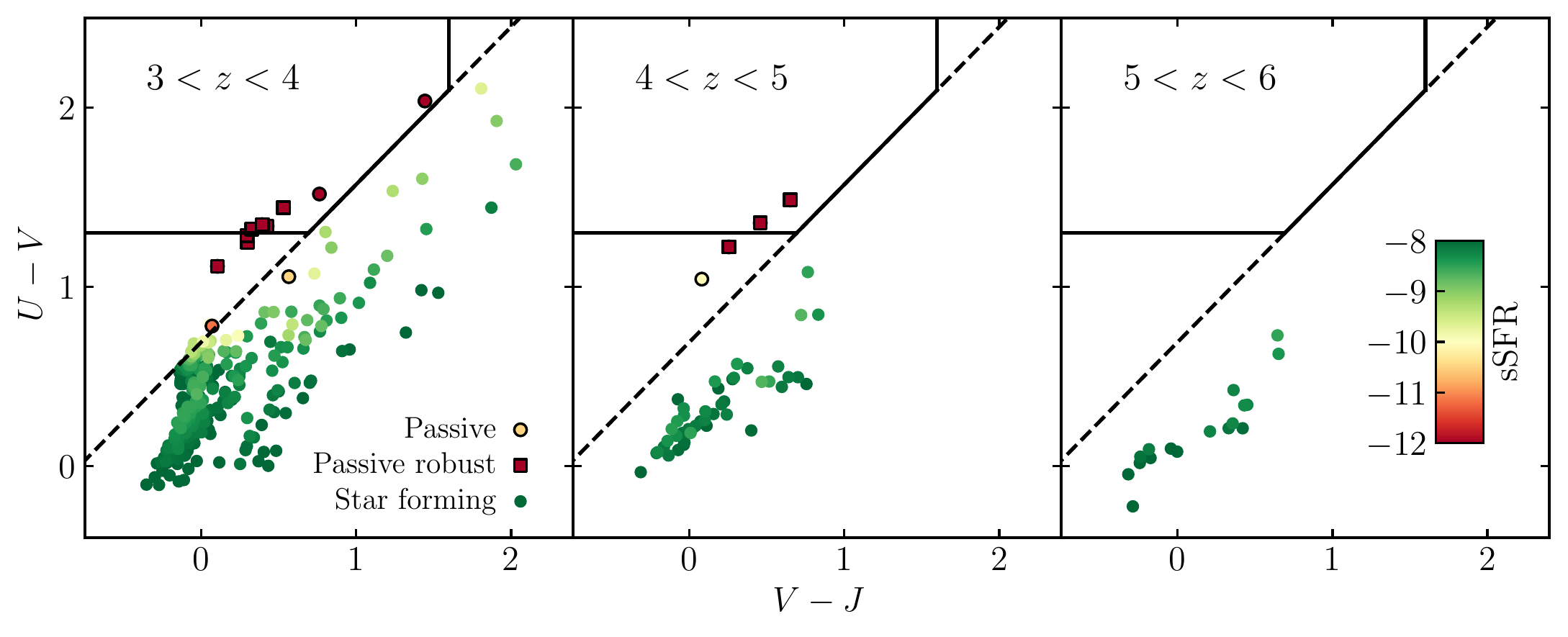}
    \caption{The rest-frame UVJ colour diagram, showing our magnitude-selected sample of massive galaxies (F200W < 26.5) in integer redshift bins spanning $3 < z < 6$. Points are coloured by sSFR, with star-forming galaxies denoted by circles without a border. Objects in our robust quiescent sub-sample are shown with black-bordered squares, whereas objects in our quiescent sample that do not meet our robust criteria (see Section \ref{method:sel}) are shown as black-bordered circles. SEDs for the three robust quiescent galaxies we identify at $z>4$ are shown in Fig. \ref{fig:spectra}, whereas the 7 robust quiescent galaxies at $3<z<4$ are shown in Fig. \ref{fig:z3_seds}.}
    \label{fig:uvj}
\end{figure*}

\section{Method}\label{method}

\subsection{Spectral energy distribution fitting}\label{method:sed}

The SED-fitting analysis in this work makes use of the \textsc{Bagpipes} spectral fitting code \citep{Carnall2018}, and is based on the method used to search for $2 < z < 5$ massive quiescent galaxies in CANDELS UDS and GOODS South in \cite{Carnall2020}.

The model we fit to our photometric data makes use of the \cite{Bruzual2003} stellar population models, in particular the 2016 updated version \citep{Chevallard2016} using the MILES stellar spectral library \citep{Sanchez-Blazquez2006, Falcon-Barroso2011} and the updated stellar evolutionary tracks of \cite{Bressan2012} and \cite{Marigo2013}.

Nebular line and continuum emission are included in our model using an approach based on the \textsc{Cloudy} photoionization code, outlined in section 3 of \cite{Carnall2018}, following \cite{Byler2017}. We assume an ionization parameter, $U=10^{-3}$, and a lifetime for stellar birth clouds of 10 Myr.

Dust attenuation is included using the model of \cite{Salim2018}, which has a variable slope, parameterised with a power-law deviation, $\delta$, from the \cite{Calzetti2000} model. We allow the $V-$band attenuation, $A_V$, to vary from $0-8$ magnitudes. We further assume that light from stars still enclosed in stellar birth clouds and resulting nebular emission is attenuated by twice the $A_V$ experienced by older stars within the wider interstellar medium (ISM) of the galaxy (e.g. \citealt{Charlot2000}).

We assume a double-power-law SFH model, as introduced in \cite{Carnall2018, Carnall2019a}, which has been shown to reproduce well the SFHs of massive quiescent galaxies in the \textsc{Mufasa} simulation \citep{Dave2016}. The stellar and nebular metallicities of galaxies are assumed to be identical, and are varied with a uniform prior in logarithmic space from $-0.7 < $ log$_{10}(Z/$Z$_\odot) < 0.4$. Intergalactic medium absorption is included using the model of \cite{Inoue2014}. We vary redshift in our model with a uniform prior over the redshift range $z=0-20$. A full list of the 9 free parameters of our model and their associated prior distributions is given in Table \ref{table:param}. We fit our \textsc{Bagpipes} model to the data using the \textsc{MultiNest} nested sampling algorithm \citep{Skilling2006, Feroz2019}, accessed via the \textsc{PyMultiNest} interface \citep{Buchner2014}.

\subsection{Selection of massive quiescent galaxies}\label{method:sel}

We begin our selection process by requiring that objects have a posterior median redshift greater than $z=3$. We also require that 97.5 per cent of the redshift posterior for each object lies above $z=2.75$. This is in order to exclude objects with significant secondary low-redshift solutions, whilst retaining objects with narrow redshift posteriors that extend marginally below $z=3$.

The sample is then cleaned by visual inspection of all 10 photometric bands, as well as the fitted \textsc{Bagpipes} SEDs. We exclude objects that fall close to the edges of the NIRCam detector, objects for which coverage is only available in some bands, and various kinds of NIRCam detector artefacts \citep{Rigby2022}. We further exclude objects that are visible in short-wavelength imaging, below the position of the Lyman break at the \textsc{Bagpipes} fitted redshift (which strongly implies the fitted redshift is incorrect). At the end of this process, we have a total of 421 galaxies at $z>3$.

To separate star-forming and quiescent galaxies, we use a time-dependent cut in specific star-formation rate (sSFR), as has been widely applied in the literature (e.g. \citealt{Gallazzi2014, Pacifici2016}). We define quiescent galaxies as those that have

\begin{equation}\label{eqn:ssfr}
\mathrm{sSFR} < \dfrac{0.2}{t_\mathrm{obs}}
\end{equation}

\noindent where $t_{\rm obs}$ is the age of the Universe at the redshift of the galaxy. This threshold is broadly equivalent to a selection in rest-frame UVJ colour space of $U\ -\ V\ >\ 0.88\ \times\ (V\ -\ J)\ +\ 0.69$ \citep{Carnall2018, Carnall2019a} at all redshifts, which is the $z<0.5$ quiescent galaxy selection criterion introduced by \cite{Williams2009}.

Our full passive sample is defined as those for which the 50th percentile of the fitted \textsc{Bagpipes} sSFR posterior distribution falls below this threshold. These objects are robustly placed at $z>3$ by our fitting, and are more likely to be quiescent than star forming. However, we cannot confidently exclude star-forming solutions in all cases. Following \cite{Carnall2020}, we then further define a ``robust'' quiescent sub-sample, for which 97.5 per cent of the sSFR posterior is required to fall below the threshold in Equation \ref{eqn:ssfr}. For these robust objects, we exclude both low-redshift and star-forming solutions with high confidence. 

By this process we identify a total of 15 objects, 10 of which satisfy our robust selection criteria. We visually inspect the available $Spitzer$-MIPS 24$\mu$m data \citep{Dickinson2007} for these objects to check for anomalously strong detections that would clearly indicate these are lower-redshift dusty star-forming galaxies, however we do not find any obvious bright detections.

\begin{table*}
  \caption{Properties of the 15 massive quiescent galaxies at $z>3$ we identify in this work. We define $t_\mathrm{form}$ as the age of the Universe corresponding to the mass-weighted age of each galaxy, and $z_\mathrm{form}$ as the corresponding redshift. For objects labelled as ``robust'', we can confidently exclude star-forming solutions, whereas for non-robust objects the posterior median solution is quiescent, but we cannot confidently exclude star-forming solutions. Additional columns for this table, including spectral fitting results from \textsc{Eazy} and \textsc{LePhare} (see Section \ref{method:extra}) are provided as supplementary online material.}
\label{table:objects}
\begingroup
\setlength{\tabcolsep}{5pt} 
\renewcommand{\arraystretch}{1.5} 
\begin{tabular}{lccccccccccl}
\hline
ID & RA & DEC & F150W & F200W & Redshift & $t_\mathrm{form}$ / Gyr & $z_\mathrm{form}$ & log$_{10}(M_*/$M$_\odot)$ & $U - V$ & $V - J$ & Robust \\
\hline
17318&214.808153&52.832201&27.33&25.73&$4.50^{+0.13}_{-0.10}$&$0.5^{+0.3}_{-0.2}$&$9.5^{+6.2}_{-2.9}$&$10.13^{+0.04}_{-0.04}$&$1.22^{+0.13}_{-0.09}$&$0.26^{+0.14}_{-0.15}$&True \\
28316&214.871228&52.845073&23.44&22.22&$3.53^{+0.12}_{-0.12}$&$1.6^{+0.1}_{-0.1}$&$3.8^{+0.2}_{-0.1}$&$10.84^{+0.04}_{-0.04}$&$1.06^{+0.03}_{-0.04}$&$0.57^{+0.12}_{-0.12}$&False \\
29497&214.760622&52.845322&22.95&21.57&$3.25^{+0.08}_{-0.08}$&$1.2^{+0.2}_{-0.3}$&$4.9^{+1.4}_{-0.6}$&$11.34^{+0.06}_{-0.06}$&$1.44^{+0.10}_{-0.06}$&$0.53^{+0.10}_{-0.08}$&True \\
36262&214.895614&52.856497&22.77&21.68&$3.26^{+0.09}_{-0.10}$&$1.2^{+0.2}_{-0.3}$&$4.8^{+1.2}_{-0.5}$&$11.06^{+0.12}_{-0.07}$&$1.11^{+0.10}_{-0.04}$&$0.11^{+0.09}_{-0.08}$&True \\
40015&214.853899&52.861358&25.25&23.23&$3.68^{+0.16}_{-0.20}$&$1.1^{+0.2}_{-0.2}$&$5.2^{+1.1}_{-0.6}$&$11.53^{+0.10}_{-0.10}$&$2.04^{+0.08}_{-0.08}$&$1.44^{+0.16}_{-0.15}$&False \\
42128&214.850568&52.866030&25.35&23.81&$4.19^{+0.12}_{-0.12}$&$0.4^{+0.3}_{-0.2}$&$10.8^{+8.2}_{-3.4}$&$11.06^{+0.07}_{-0.05}$&$1.48^{+0.14}_{-0.07}$&$0.65^{+0.13}_{-0.12}$&True \\
44362&214.879163&52.869187&25.32&24.48&$3.39^{+0.29}_{-0.14}$&$1.7^{+0.1}_{-0.2}$&$3.7^{+0.3}_{-0.1}$&$9.63^{+0.04}_{-0.03}$&$0.78^{+0.03}_{-0.03}$&$0.07^{+0.09}_{-0.08}$&False \\
52124&214.866027&52.884091&24.7&23.23&$3.38^{+0.19}_{-0.09}$&$1.2^{+0.3}_{-0.3}$&$4.8^{+1.2}_{-0.8}$&$10.81^{+0.11}_{-0.05}$&$1.52^{+0.05}_{-0.09}$&$0.76^{+0.14}_{-0.17}$&False \\
52175&214.866039&52.884255&23.84&22.31&$3.44^{+0.14}_{-0.08}$&$1.2^{+0.1}_{-0.2}$&$4.8^{+0.7}_{-0.3}$&$10.87^{+0.05}_{-0.02}$&$1.32^{+0.04}_{-0.10}$&$0.33^{+0.05}_{-0.10}$&True \\
75768&214.904841&52.935352&24.51&23.25&$3.31^{+0.10}_{-0.07}$&$1.3^{+0.2}_{-0.3}$&$4.6^{+1.2}_{-0.4}$&$10.48^{+0.07}_{-0.04}$&$1.25^{+0.07}_{-0.06}$&$0.30^{+0.08}_{-0.09}$&True \\
80785&214.915559&52.949026&24.76&23.96&$4.86^{+0.15}_{-0.16}$&$0.5^{+0.2}_{-0.2}$&$9.2^{+3.6}_{-2.1}$&$10.90^{+0.08}_{-0.05}$&$1.04^{+0.09}_{-0.04}$&$0.08^{+0.18}_{-0.08}$&False \\
8888&214.767258&52.817698&25.19&23.73&$3.49^{+0.17}_{-0.12}$&$1.2^{+0.2}_{-0.3}$&$4.9^{+1.1}_{-0.6}$&$10.54^{+0.12}_{-0.08}$&$1.34^{+0.06}_{-0.06}$&$0.43^{+0.12}_{-0.10}$&True \\
92564&214.957874&52.980293&24.9&23.35&$3.47^{+0.11}_{-0.10}$&$1.2^{+0.1}_{-0.1}$&$4.9^{+0.5}_{-0.4}$&$10.49^{+0.08}_{-0.04}$&$1.29^{+0.07}_{-0.07}$&$0.30^{+0.09}_{-0.11}$&True \\
97581&214.981800&52.991238&24.14&22.64&$3.46^{+0.10}_{-0.09}$&$1.0^{+0.2}_{-0.2}$&$5.5^{+1.0}_{-0.8}$&$10.81^{+0.06}_{-0.03}$&$1.35^{+0.07}_{-0.10}$&$0.40^{+0.05}_{-0.11}$&True \\
101962&215.039054&53.002778&26.39&24.85&$4.39^{+0.11}_{-0.14}$&$0.4^{+0.3}_{-0.2}$&$12.1^{+6.3}_{-4.5}$&$10.63^{+0.06}_{-0.04}$&$1.36^{+0.11}_{-0.07}$&$0.46^{+0.16}_{-0.11}$&True \\
\hline
\end{tabular}
\endgroup
\end{table*}

\subsection{Comparison of our results with other codes}\label{method:extra}

As an additional check on our results, we fit our photometry for these 15 objects with two additional SED fitting codes. We firstly run \textsc{Eazy} \citep{Brammer2008}, using both the Pegase and principal component analysis (PCA) template sets, and additionally run \textsc{LePhare} \citep{Arnouts1999, Ilbert2006}, using the \cite{Bruzual2003} models. The redshifts for these three spectral fitting runs, along with best-fit masses and sSFRs from \textsc{LePhare} are provided in the online version of Table \ref{table:objects} as supplementary material.

For our robust sub-sample, all three additional sets of redshifts are in good agreement with our \textsc{Bagpipes} results, with typical variations of d$z \simeq 0.1-0.2$. The masses returned by \textsc{LePhare} are also very similar to our \textsc{Bagpipes} posterior median values, with a mean offset of 0.07 dex. \textsc{LePhare} also returns sSFRs below the threshold defined in Equation \ref{eqn:ssfr} for all 10 objects in our robust sub-sample.

For our 5 non-robust objects, the agreement is similar for 4 objects, however for the remaining object (ID: 44362), \textsc{LePhare} finds a lower-mass star-forming solution, though all of the codes still find this object to be at $z\simeq3$. This finding reflects the lower level of certainty we attach to objects that do not meet our robust selection criteria.

\section{Results}\label{results}

From our analysis, we identify a total of 15 quiescent galaxies at $z>3$, of which 10 are members of our robust sub-sample. From these 15 objects, 4 are placed reliably at $z>4$, and 3 of these are robustly identified as quiescent. This is the first clear identification of massive quiescent galaxies significantly beyond $z=4$. We provide coordinates, photometric redshifts, magnitudes and physical properties for the 15 quiescent galaxies we identify in Table \ref{table:objects}. We present SEDs for our 10 robust objects in Figs \ref{fig:spectra} and \ref{fig:z3_seds}. Cutout images for these 10 robust galaxies are presented in Fig. \ref{fig:cutouts}.

\subsection{Robust massive quiescent galaxies at $\mathbf{z>4}$}

SEDs and colour images for the three $z>4$ objects in our robust sub-sample are shown in Fig. \ref{fig:spectra}. All three of these display a strong Balmer break between the F200W and F277W bands, which provides a very strong constraint on their redshifts. They also exhibit well-constrained red spectral slopes in the rest-frame near-UV, indicating a lack of ongoing star formation, and blue spectral slopes in the longer-wavelength NIRCam bands, strongly ruling our lower-redshift dusty solutions. In this section we briefly discuss the observed properties of each object, before moving on to discuss their SFHs in Section \ref{discussion}.

\subsubsection{Galaxy 17318}

The highest-redshift robust quiescent galaxy in our sample is object 17318, with F200W = 25.7 and a photometric redshift of $z~\simeq~4.5$. This galaxy, as shown in the top panel of Fig. \ref{fig:spectra}, exhibits the characteristic triangular SED shape of post-starburst (PSB) galaxies at lower redshift (e.g. \citealt{Wild2014}), implying a recent, rapid fall in SFR (e.g. \citealt{Wild2020, Deugenio2021}). This is perhaps unsurprising, considering this object is observed just $\simeq1.4$ billion years after the Big Bang. For this galaxy we derive a stellar mass of log$_{10}(M_*/$M$_\odot) = 10.13\pm0.06$, and a $2\sigma$ upper limiting sSFR (97.5th percentile) of log$_{10}($sSFR/yr$^{-1}) = -10.6$.

Our sample is shown on the rest-frame UVJ colour diagram in Fig. \ref{fig:uvj}. Galaxy 17318 is the bluest (closest to the bottom-left) robust object shown in the central panel, just below the horizontal edge of the solid UVJ selection box, again highly consistent with lower-redshift PSBs (e.g. \citealt{Belli2019, Carnall2019b}). The locations of $z > 3$ quiescent galaxies on various other rest-frame colour selection diagrams (e.g., NUVrJ; \citealt{Ilbert2013}) will be explored in upcoming work by Gould et al. in prep.

This object is at a similar redshift to GOODSS-9209, the highest-redshift candidate identified by \cite{Carnall2020}, which is the target of NIRSpec Cycle 1 observations \citep{Carnall2023c}. This new galaxy however is $\simeq0.6$ dex less massive, and approximately 2 magnitudes fainter at $\lambda=2\,\mu$m (GOODSS-9209 has $K_{s}=23.6$).

This object is not included in the CANDELS catalogue of \cite{Stefanon2017}. We measure a F150W magnitude for this object of 27.4, which is fainter than the 50 per cent point-source completeness limit calculated by \cite{Stefanon2017} in their F160W selection band of 27.23. It is therefore not surprising that this object was not included.

\subsubsection{Galaxy 101962}\label{sec:101962}

The second-highest-redshift robust candidate we identify is object 101692 at $z~\simeq~4.4$, shown in the middle panel of Fig. \ref{fig:spectra}. It is slightly redder than 17318, and is the middle robust quiescent object shown on the UVJ diagram in the central panel of Fig. \ref{fig:uvj}. In most respects however, this object is quite similar to our other $z > 4$ robust candidates. For this galaxy we derive a stellar mass of log$_{10}(M_*/$M$_\odot) = 10.63^{+0.06}_{-0.04}$ (approximately 0.5 dex more massive than 17318) and a $2\sigma$ upper limiting sSFR of log$_{10}($sSFR/yr$^{-1}) = -9.9$.

This is the only one of our three robust $z>4$ candidates to appear in the CANDELS EGS catalogue of \cite{Stefanon2017}, having a match within 0.1$^{\prime\prime}$ (ID: 23297). We show the CANDELS photometry for this object in blue in Fig. \ref{fig:spectra2}. This galaxy was only previously detected with $3\sigma$ significance in the WFCAM $K_{s}$ band, and no data were available at all between $\lambda=2.3-3.0\mu$m, meaning virtually no constraint could be placed on the Balmer break. The median photometric redshift reported by the CANDELS team (following the method of \citealt{Dahlen2013}) is $z=3.5$. However, the individual estimates from their different codes display considerable variance, with a $1\sigma$ range from $z=2.45-4.70$. 

As a demonstration, we fit the \cite{Stefanon2017} photometry shown in Fig. \ref{fig:spectra2} with \textsc{Bagpipes}, using the same model described in Section \ref{method:sed}. Our best fit is shown in green in Fig. \ref{fig:spectra2}, along with the gold JWST data and red best fit from Fig. \ref{fig:spectra}. From the \cite{Stefanon2017} data we recover an extremely broad photometric redshift posterior, $z=1.34^{+1.94}_{-0.39}$. The 1D and 2D stellar mass, sSFR and redshift posteriors for both fits are shown in the corner plot on the left side of Fig. \ref{fig:spectra2}. It can be seen that the lack of a well-constrained redshift results in a very broad stellar mass posterior, and virtually no constraint on the sSFR, with the sSFR posterior following the prior imposed by our SFH model (e.g., see Fig. 1 of \citealt{Carnall2019a}).

\subsubsection{Galaxy 42128}

Galaxy 42128 has a slightly lower redshift, $z\simeq4.2$, and is both brighter and more massive than the other two galaxies. It is also the reddest of our $z > 4$ candidates on the UVJ diagram. However, in most respects, this galaxy is very similar to 17318 and 101962, also exhibiting the characteristic triangular PSB spectral shape. For this galaxy, we derive a stellar mass of log$_{10}(M_*/$M$_\odot) = 11.06^{+0.07}_{-0.05}$, and a $2\sigma$ upper limiting sSFR of log$_{10}($sSFR/yr$^{-1}) = -11.2$.

In shorter-wavelength, lower spatial resolution imaging this 
object is not distinguishable from the extended structure of the nearby barred-spiral galaxy and therefore does not feature in the \cite{Stefanon2017} CANDELS EGS catalogue. However, the longer wavelength and higher spatial resolution imaging provided by JWST reveal 
this object as a brighter, redder, highly compact background source.

\begin{table}
  \caption{Number densities derived for our quiescent galaxy sample in integer redshift bins spanning $3 < z < 5$. Uncertainties were calculated as the Poisson noise on the number of objects found \protect\citep{Gehrels1986}. The full quiescent sample includes quiescent galaxies for which we cannot rule out secondary star-forming solutions, whereas the robust sub-sample includes only those galaxies for which we can confidently exclude star-forming solutions.}
\label{table:ndensity}
\begingroup
\setlength{\tabcolsep}{6pt} 
\renewcommand{\arraystretch}{1.25} 
\begin{tabular}{cccc}
\hline
\multicolumn{4}{c}{Full quiescent sample} \\
\hline
& $N_\mathrm{galaxies}$ & $n$ / Mpc$^{-3}$ & $n$ / Mpc$^{-3}$ \\
Redshift range & F200W < 26.5 & F200W < 26.5 & F200W < 24.5 \\
\hline
$3 < z < 4$ & 11 & $11.6^{+4.7}_{-3.4}\times10^{-5}$
 & $10.6^{+4.5}_{-3.3}\times10^{-5}$ \\
$4 < z < 5$ & 4 & $4.7^{+3.7}_{-2.2}\times10^{-5}$
 & $2.3^{+3.1}_{-1.5}\times10^{-5}$ \\
\hline
\multicolumn{4}{c}{Robust sub-sample} \\
\hline
Redshift range & $N_\mathrm{galaxies}$ & $n$ / Mpc$^{-3}$ & $n$ / Mpc$^{-3}$ \\
 & F200W < 26.5 & F200W < 26.5 & F200W < 24.5 \\
\hline
$3 < z < 4$ & 7 & $7.4^{+4.0}_{-2.7}\times10^{-5}$
 & $6.3^{+3.8}_{-2.5}\times10^{-5}$ \\
$4 < z < 5$ & 3 & $3.5^{+3.4}_{-1.9}\times10^{-5}$
 & $1.2^{+2.7}_{-1.0}\times10^{-5}$ \\
\hline
\end{tabular}
\endgroup
\end{table}

\begin{figure}
	\includegraphics[width=\columnwidth]{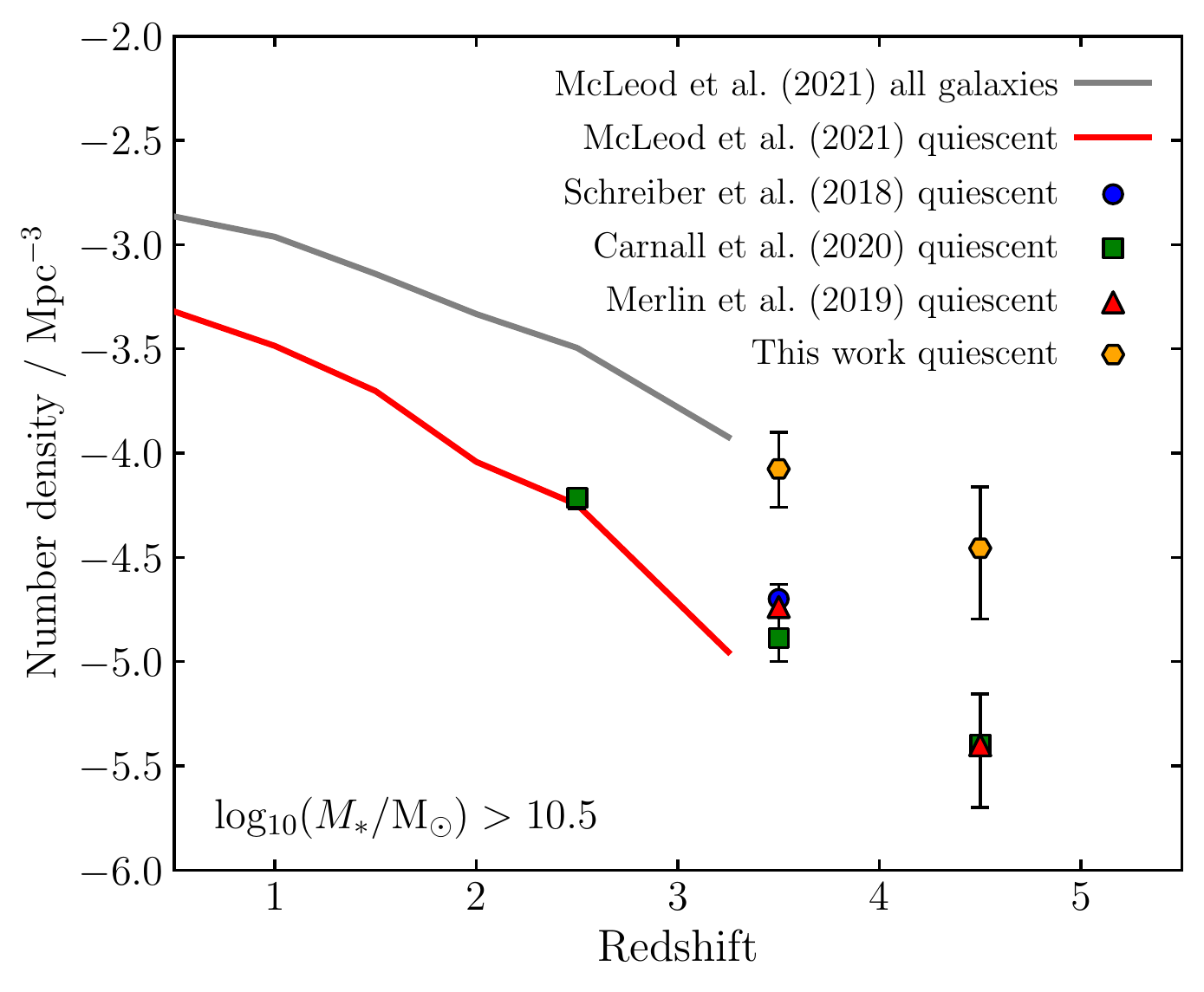}
    \caption{Number density estimates for high-redshift massive quiescent galaxies. Our estimate at $3 < z < 5$ derived from the JWST CEERS data are $3-5$ times higher than pre-JWST estimates, and, at $z\simeq3$, approach the result of \protect \cite{McLeod2021} for the total galaxy population. Stellar masses derived by other authors have been converted to a \protect \cite{Kroupa2001} IMF where necessary.}
    \label{fig:ndensity}
\end{figure}

\subsection{Comparison with previous work in the EGS field}

A previous search for $z > 3$ massive quiescent galaxies in the EGS field was presented by \cite{Merlin2019}, who analysed the CANDELS catalogue of \cite{Stefanon2017}. In this subsection we present a comparison with their results to demonstrate the power of the new JWST data. From our 15-object sample, 12 objects have matches in the \cite{Stefanon2017} catalogue, including object 101963 discussed in Section \ref{sec:101962}.

\cite{Merlin2019} identify 13 candidate $z > 3$ quiescent galaxies in the \cite{Stefanon2017} catalogue. From these, 4 objects fall within the area covered by our JWST CEERS catalogue (a further object falls within a gap between two NIRCam short-wavelength detectors). All 4 of these are also included in our sample (IDs: 28316, 36262, 52175, 75768). These are all at $z\lesssim 3.5$, and are the 4 brightest galaxies in our sample in the F150W band. The other 8 objects in our sample that also appear in the \cite{Stefanon2017} catalogue were not selected by \cite{Merlin2019}, likely as a result of their relative faintness, which makes constraining redshifts, masses and sSFRs more challenging, as demonstrated in Fig. \ref{fig:spectra2}.

\subsection{Quiescent galaxy number densities at $\mathbf{3 < z < 5}$}

In Table \ref{table:ndensity} we report our estimates of the number densities of quiescent galaxies over the redshift range $3 < z < 5$ that meet our F200W < 26.5 selection threshold. We report number densities based on both our full quiescent sample and our robust sub-sample. The numbers based on the robust sub-sample can be interpreted as conservative lower limits.

In \cite{Schreiber2018}, the authors report a number density of $2.0\pm0.3\ \times\ 10^{-5}$ Mpc$^{-3}$ for a spectroscopic sample of quiescent galaxies at $3 < z < 4$ with $K_{s} < 24.5$. We compare our results with \cite{Schreiber2018} by calculating the number of F200W < 24.5 quiescent galaxies in our sample at $3 < z < 4$, as this filter is closest in wavelength coverage to the $K_{s}$ band. These numbers are reported in the right-hand column of Table \ref{table:ndensity}. 

The number density we find for our robust sub-sample at F200W < 24.5 is approximately a factor of 3 larger than the result of \cite{Schreiber2018}, whereas the result from our full quiescent sample is approximately a factor of 5 larger. We attribute this significant increase in the number of $z > 3$ quiescent galaxies to the much deeper, redder imaging now available from JWST, which allows physical properties to be reliably inferred for faint, red galaxies such as these.

From our 11 quiescent galaxies at $3 < z < 4$, a total of 10 have matches within 0.25$^{\prime\prime}$ in the \cite{Stefanon2017} CANDELS EGS catalogue. From these 10, a total of 9 have CANDELS photometric redshifts in the range from $3 < z < 4$, and by fitting the \cite{Stefanon2017} photometry with \textsc{Bagpipes} we recover similar results. However, just 3 of these objects would be included in our quiescent sample given the CANDELS photometry, as their sSFRs are far less well constrained by these data. Just 2 of these 3 would be identified as robust.

It is therefore likely that, had we designed a spectroscopic follow-up campaign similar to that of \cite{Schreiber2018} based on fitting only CANDELS EGS data, we would have arrived at a very similar number density for $3 < z < 4$ quiescent galaxies to the one they obtain. Indeed, our result based on fitting CANDELS UDS and GOODS South photometry in \cite{Carnall2020} arrived at a number density of $1.7\pm0.3\ \times\ 10^{-5}$ Mpc$^{-3}$ for $3 < z < 4$ quiescent galaxies with $K_s < 24.5$, fully consistent with \cite{Schreiber2018}.

In Fig. \ref{fig:ndensity} we show a comparison of $z > 3$ quiescent galaxy number density estimates from the literature, including our new results. We restrict this comparison to galaxies with log$_{10}(M_*/\mathrm{M_\odot}) > 10.5$, where the CANDELS catalogues are mass complete at $z < 4$. At $z > 4$, the Balmer break moves to $\lambda>2\mu$m, meaning selection at longer wavelengths is likely to be necessary to obtain mass-complete samples.

As above, it can be seen from Fig. \ref{fig:ndensity} that our new result (for our full sample) is a factor of $\simeq5$ higher than previous estimates, which are all in close agreement. We also show results at $z < 3.75$ derived from the stellar mass functions of \cite{McLeod2021}, both for quiescent galaxies and for the whole galaxy population. Our $3 < z < 4$ quiescent galaxy number density is close to the total number density found by \cite{McLeod2021}. This suggests the small area studied in this work may be over-dense at these redshifts.

Finally we note that, as shown in \cite{Carnall2019a}, standard parametric SFH models impose strong priors on galaxy sSFRs, favouring sSFRs close to the star-forming main sequence. In the absence of strongly constraining data, this could plausibly result in quiescent galaxies being misidentified as star forming. This effect is demonstrated in Fig. \ref{fig:spectra2}, in which the sSFR posterior for galaxy 101962 using CANDELS data is strongly weighted towards star-forming solutions, even at the correct redshift of $z\simeq4.4$. Clearly, follow-up studies using larger-area JWST imaging surveys are of critical importance to clarify this situation.

\begin{figure}
	\includegraphics[width=\columnwidth]{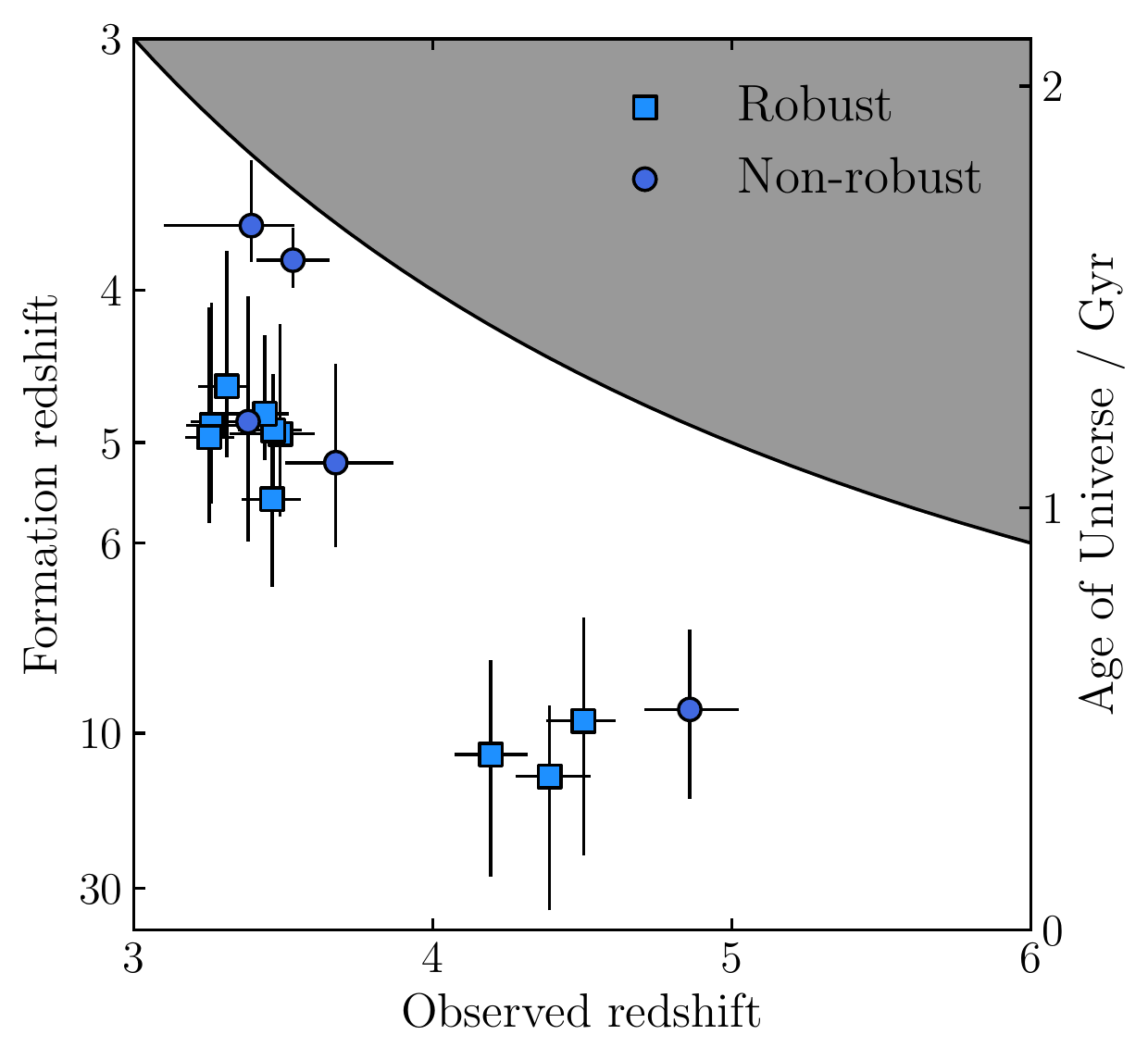}
    \caption{Formation redshifts for $z>3$ massive quiescent galaxy sample, shown as a function of their observed redshifts. Our candidates at $z>4$ have formation redshifts from $9 < z_\mathrm{form} < 12$, whereas our $3 < z < 4$ galaxies are all younger, having formed at $z_\mathrm{form} < 6$.}
    \label{fig:tform}
\end{figure}

\begin{figure}
	\includegraphics[width=\columnwidth]{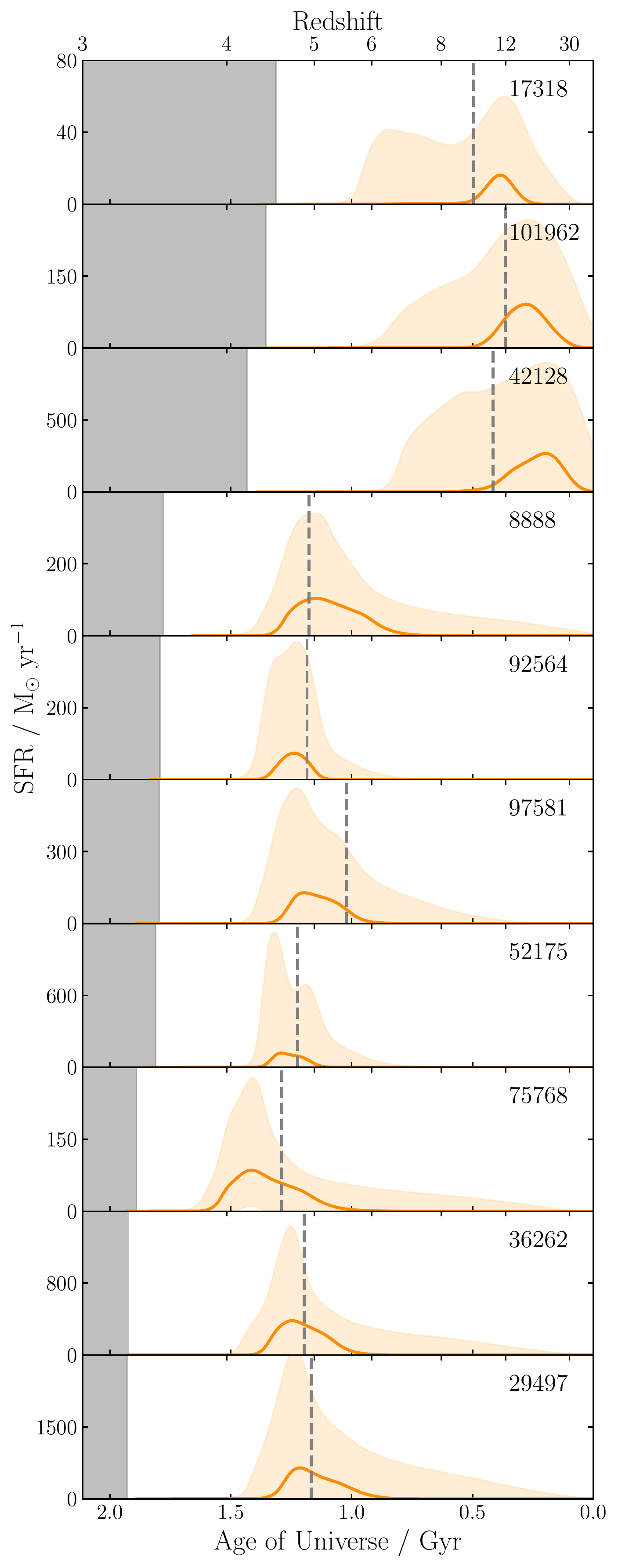}
    \caption{Star-formation histories for our 10 robustly identified $z>3$ quiescent galaxies. The solid orange lines show the 50th percentiles of the SFH posterior distributions in each case, whereas the orange shaded regions show the $1\sigma$ range. The dashed gray vertical lines show the posterior median redshift of formation for each object. SEDs are shown in Figs \ref{fig:spectra} and \ref{fig:z3_seds}.}
    \label{fig:sfhs}
\end{figure}

\section{Discussion}\label{discussion}

In this section we discuss the star-formation histories we infer for our sample. We calculate formation times, $t_\mathrm{form}$, for each object as the average time at which the stars in the galaxy formed. This is the age of the Universe at the time corresponding to the (mass-weighted) mean stellar age (see \citealt{Carnall2018}, Equation 11). We also calculate formation redshifts, $z_\mathrm{form}$, which are the redshifts corresponding to $t_\mathrm{form}$. These values are reported in Table \ref{table:objects}. 

Formation redshifts for our sample are plotted against the observed redshift of each galaxy in Fig. \ref{fig:tform}. In addition, the full SFH posteriors for the 10 galaxies in our robust sub-sample are shown in Fig. \ref{fig:sfhs}. Their posterior median formation redshifts are shown with dashed gray vertical lines. 

It can be seen that the 3 robust quiescent galaxies at $z>4$ formed the bulk of their stellar populations during the first billion years at $z>6$, with posterior median formation redshifts in the range $9 < z_\mathrm{form} < 12$. The second-highest-redshift object, 101962, is the earliest formed, with the bulk of its stars having formed at $z > 8$. These findings make our $z>4$ objects highly plausible as descendants of the sample identified by \cite{Labbe2022}. Indeed, objects 101962 and 42128 are both predicted to have had log$_{10}(M_*/$M$_\odot)>10$ at their formation redshifts of $z_\mathrm{form}=12.1^{+6.3}_{-4.5}$ and $z_\mathrm{form}=10.8^{+8.2}_{-3.4}$ respectively. This is fully consistent with the finding of \cite{Labbe2022} that a considerable number of log$_{10}(M_*/$M$_\odot)>10$ galaxies were already in place by $7 < z < 11$.
 
The 7 robust quiescent galaxies at $3 < z < 4$ formed their stellar populations later in cosmic time, with $4 < z_\mathrm{form} < 6$. The fact that none of these galaxies is older suggests that the $z>4$ quiescent galaxies in our sample have not yet reached the end-point of their evolution, and are likely to experience further star formation by $z\simeq3$. However, larger-area JWST surveys and spectroscopic follow-up will certainly be required to rule out the possibility of $3 < z < 4$ quiescent galaxies with stellar populations dating back to $z>6$.

It is interesting to reflect on the fact that we do not identify any quiescent galaxies at $z>5$, despite the fact that our oldest object appears to have quenched by $z>6$. Apart from the obvious limitations of the relatively small imaging area included in the initial CEERS release, it is currently unclear how the colours of newly quenched galaxies evolve in order to arrive in the UVJ-quiescent box, and how long this might take after star formation ceases (e.g. \citealt{Belli2019, Carnall2019a, Akins2022}).

Upcoming larger-area JWST imaging surveys, such as Public Release Imaging for Extragalactic Research (PRIMER\footnote{\url{https://primer-jwst.github.io}}) are ideally suited to searching for evidence of quiescent galaxies at $z>5$, as well as selecting larger samples of $3 < z < 5$ quiescent galaxies to produce more-robust number densities. Such searches may benefit from the use of catalogues selected in longer-wavelength NIRCam bands, rather than the F200W selection employed in this work, as well as new, JWST-specific colour selection criteria (e.g. \citealt{Leja2019b, Antwi-Danso2022}).

\section{Conclusion}\label{conclusion}

In this work we present the results of a search for massive quiescent galaxies at redshifts $z > 3$, selected from the first NIRCam data taken by the JWST CEERS Early Release Science programme. We identify 15 galaxies in the redshift range $3 < z < 5$ with robust photometric redshifts and posterior median sSFRs that suggest they are quiescent. For 10 of these galaxies, we can confidently rule out star-forming solutions, and we refer to these galaxies as comprising our robust sub-sample.

Three of our robustly quiescent objects are at $z>4$, and these constitute the best evidence to date for the existence of massive quiescent galaxies significantly above $z=4$. These objects have stellar masses in the range  log$_{10}(M_*/$M$_\odot)=10.1-11.1$ and posterior median formation redshifts from $9 < z_\mathrm{form} < 12$. Two of these $z>4$ galaxies would have had stellar masses in excess of log$_{10}(M_*/$M$_\odot)=10$ by $z\gtrsim8$, supporting the recent findings of \cite{Labbe2022}.

Only one of these $z>4$ objects has a match in the \cite{Stefanon2017} CANDELS EGS catalogue. The redshift of this object was not well constrained by previously available data, with both the CANDELS team and \textsc{Bagpipes} returning extremely broad redshift posteriors, with best-fitting values at $z < 4$.

We calculate number densities for our quiescent sample, as well as for our robust sub-sample (which can be regarded as yielding robust lower limits). We find that the number density of $3 < z < 4$ quiescent galaxies with F200W < 24.5 is a factor of $3-5$ times higher than previously reported by \cite{Schreiber2018}. We demonstrate that this difference arises as a result of better constraints from the new NIRCam data, and show that previously available data from CANDELS would have led us to a similar number density to that calculated by \cite{Schreiber2018}. This finding poses an additional challenge for simulations of early galaxy formation, which already struggle to reproduce previously reported number densities.

The 10 robust $z>3$ massive quiescent galaxies we report, and the 3 at $z>4$ in particular, are excellent potential targets for follow up NIRSpec spectroscopy, either as part of CEERS, or via dedicated programmes in JWST Cycle 2 and beyond. The detection of Balmer absorption features would unambiguously confirm these objects as quiescent at $z>4$, and full spectral fitting of deep continuum spectroscopic data would provide strong constraints on their SFHs, as well as the ability to probe in detail their physical properties.

\section*{Acknowledgements}

A.\,C. Carnall thanks the Leverhulme Trust for their support via a Leverhulme Early Career Fellowship. R. Begley, D.\,J. McLeod, M.\,L. Hamadouche, C. Donnan, R.\,J. McLure, J.\,S.~Dunlop and V. Wild acknowledge the support of the Science and Technology Facilities Council. F. Cullen acknowledges support from a UKRI Frontier Research Guarantee Grant [grant reference EP/X021025/1]. S. Jewell and C. Pollock acknowledge the support of the School of Physics \& Astronomy, University of Edinburgh via Summer Studentship bursaries.

\section*{Data Availability}

All JWST and HST data products are available via the Mikulski Archive for Space Telescopes (\url{https://mast.stsci.edu}). Photometric data and fitted model posteriors are available upon request.

\bibliographystyle{mnras}
\bibliography{carnall2023b}

\appendix

\section{SEDs and cutout images for robust quiescent galaxies}

\begin{figure*}
	\includegraphics[width=\textwidth]{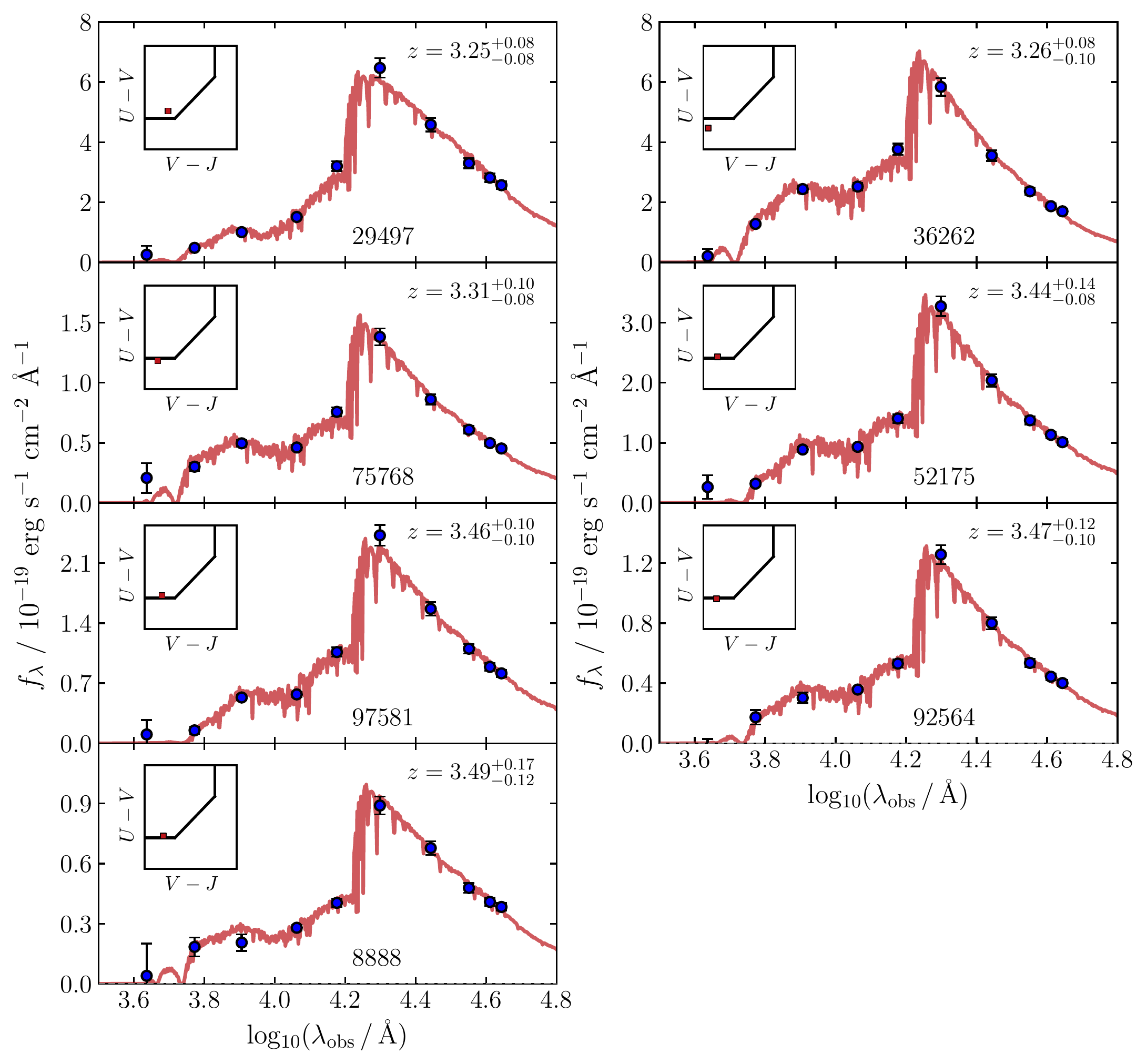}
    \caption{Spectral energy distributions for our 7 robust $3 < z < 4$ quiescent galaxies (SEDs for the three robust galaxies at $z>4$ are shown in Fig. \ref{fig:spectra}). Our 10-band photometric data from HST ACS and JWST NIRCam are shown in blue. The posterior median \textsc{Bagpipes} models are overlaid in red. The inset panels show the position of each object on the UVJ diagram.}
    \label{fig:z3_seds}
\end{figure*}

\begin{figure*}
	\includegraphics[width=\textwidth]{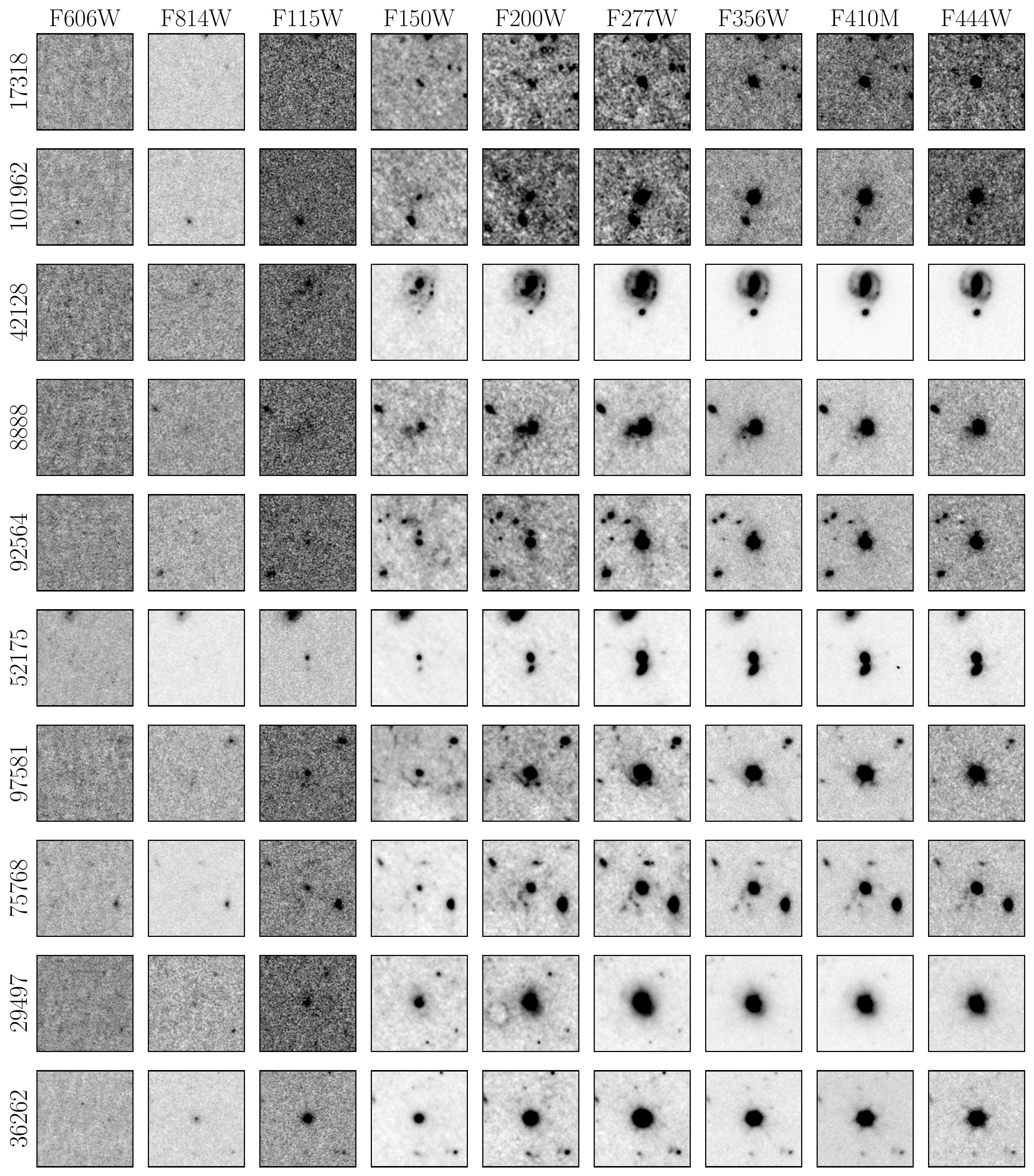}
    \caption{Cutout images for each of the 10 objects in our robust quiescent sample, in descending order of redshift. Each cutout image is $5^{\prime\prime}\ \times\ 5^{\prime\prime}$.}
    \label{fig:cutouts}
\end{figure*}

In Fig. \ref{fig:z3_seds} we show SEDs for the 7 robust quiescent galaxies we identify at $3 < z < 4$. SEDs for the 3 robust quiescent galaxies we identify at $z>4$ are shown in Fig. \ref{fig:spectra}. In Fig. \ref{fig:cutouts} we show $5^{\prime\prime}\ \times\ 5^{\prime\prime}$ HST ACS + JWST NIRCam cutout images for each of the 10 galaxies in our robust sample.

\label{lastpage}
\end{document}